\documentclass{article}

\usepackage{arxiv}

\usepackage[utf8]{inputenc} 
\usepackage[T1]{fontenc}    
\usepackage{hyperref}       
\usepackage{url}            
\usepackage{booktabs}       
\usepackage{amsfonts}       
\usepackage{nicefrac}       
\usepackage{microtype}      
\usepackage{lipsum}
\usepackage{graphicx}
\usepackage{tabularx}
\usepackage{multirow}
\graphicspath{ {./images/} }

\title{Network Learning approaches to study World Happiness}

\author{
 Meghna Chaudhary$^{\dagger}$ \thanks{ $^{\dagger}$The authors contribute equally to this paper} \\
  Department of Computer Science\\
  Shiv Nadar University\\
  Uttar Pradesh, India\\
  \texttt{mc958@snu.edu.in} \\
   \And
 Siddharth Dixit $^{\dagger  *}$\\
  Department of Mathematics\\
  Shiv Nadar University\\
  Uttar Pradesh, India\\
  \texttt{sd882@snu.edu.in} \\
  \And
  Niteesh Sahni \\
  Department of Mathematics\\
  Shiv Nadar University\\
  Uttar Pradesh, India\\
  \texttt{niteesh.sahni@snu.edu.in} \\
}

\begin{document}
\maketitle
\begin{abstract}
The United Nations in its 2011 resolution declared the pursuit of happiness a fundamental human goal and proposed public and economic policies centered around happiness. In this paper we used 2 types of computational strategies viz. \textit{Predictive Modelling} and \textit{Bayesian Networks (BNs)} to model the processed historical happiness index data of 156 nations published by UN since 2012.  We attacked the problem of prediction using General Regression Neural Networks (GRNNs) and show that it out performs other state of the art predictive models. To understand causal links amongst key features that have been proven to have a significant impact on world happiness, we first used a manual discretization scheme to discretize continuous variables into  3 levels viz. \textit{Low, Medium} and \textit{High}. A consensus World Happiness BN structure was then fixed after amalgamating information by learning 10000 different BNs using bootstrapping. Lastly, exact inference through conditional probability queries was used on this BN to unravel interesting relationships among the important features affecting happiness which would be useful in policy making.
\end{abstract}

\keywords{World Happiness \and General Regression Neural Networks \and Bayesian Networks \and Policy Making}

\section{Introduction}
The result of a resolution that was adopted by the UN in 2011 declared the pursuit of happiness “a fundamental human goal”. Member countries were invited to measure the happiness of their people and use that data towards a more holistic approach regarding public policy and economic growth. According to reports by the OECD, more than 20 countries have begun making use of subjective well-being data as part of their process of policy-making. O’Donnell et al. \cite{o2015national} had also discussed the possible use of happiness and well-being measures instead of traditional economic measures in government policy-making. This idea of looking beyond economic progress, however, is not new. Since the early 1970s, Bhutan has pursued citizen well-being over material wealth, using Gross National Happiness index (GNH) as a measure of its progress.\newline
But why does happiness matter so much? Numerous studies have shown that happiness has a wide range of benefits \cite{lyubomirsky2005benefits}. According to research, happy physicians tend to make faster and more accurate diagnosis \cite{estrada1997positive}. Not only this, happiness also enhances the learning ability of students\cite{durlak2011impact}. Apart from the aforementioned personal benefits, happier people enjoy better health, live longer lives\cite{diener2009assessing}, avoid risky behavior on the roads and are thus less likely to be involved in accidents \cite{goudie2014happiness}, and have an overall positive impact on society \cite{guven2011happier}.\newline
Owing to the importance of happiness, the United Nations Sustainable Development Solutions Network has published World Happiness Reports since 2012. These reports attempt to explain happiness using several factors including but not limited to GDP per capita, social support, healthy life expectancy at birth, perceptions of corruption and generosity. Statistical learning techniques have been used in the past to accurately model and better understand happiness and the factors that affect it. However, unlike most previous studies, rather than just looking at how and to what extent various factors can affect national happiness, in the present study we also explored how the aforementioned factors interact with each other at the same time. This approach contributes towards a better understanding of how these factors work together to promote or impede happiness in a country.

Happiness, subjective well-being and satisfaction with life have been common topics of research over the years. Diener [1984] mentioned a great need for a stronger connection to be established between the theory and research of subjective well-being. Later it was noted that measures of negative reactions like depression and anxiety could not give the complete picture of a person’s well-being \cite{diener2009factors}. Balatsky et al. \cite{balatsky1993subjective} suggested that subjective well-being was structurally invariant across different cultures. Happiness in both individual and national contexts, has been the subject of thorough research in the recent past.\newline
Correlations of a variety of economic, social, and cultural factors with happiness has been explored in detail in previous work [\cite{schyns1998crossnational}; \cite{diener2009assessing} ; \cite{cordero2017exploring}].
Apart from manual analysis of happiness and the factors affecting it, several people have also made use of machine learning techniques to examine happiness. In 2011, Binder and Coad used  cross-sectional data from the British Household Panel Survey (BHPS) from the year 2006 to analyse the effects of various factors like income, health and social factors in different quintiles of happiness. They believed that using standard regression analysis techniques focused on average values did not give the complete picture. Therefore, they applied quantile regression for a more robust exploration of happiness.\newline
Another study soon after this \cite{lang2012most} used data from the United Nations Development Programme (UNDP) Human Development Reports, The World Bank, the New Economic Foundation's The (Un) Happy Planet Index 2.0, and The World Database of Happiness, the Satisfaction with Life Scale to build 3 regression models. These models were used to determine the most influential factors affecting happiness and were based off of 3 different indices used to quantify happiness: the Happy Planet Index, the Satisfaction with Life Scale and the World Database of Happiness. The models were successful in determining the importance of corruption, Human development index, unemployment and income distribution in the context of happiness.\newline
Later, Garaigordobil [2015] conducted a study on a sample consisting of adolescents aged 14-16 years, exploring the predictive factors of happiness such as health and  sociability of an individual by conducting regression analysis. Campos et al. \cite{campos2016meditation}  also investigated the influence of factors such as mindfulness and self-compassion on the happiness of an individual using hierarchical regression analysis (along with multivariate analyses of covariance (MANCOVA)). While Saputri and Lee [2015] used data of 187 countries across the globe gathered from the UN development project to use Support Vector Machine (SVM) to predict the happiness of a country as unhappy, mid or happy. A more recent study \cite{perez2019happiness} used conceptual data structures to create a deep neural network to model happiness. They argued that most conventional methodologies, such as the commonly used Multivariate Linear Regression analysis, do not possess the capacity to truly represent complex psychological factors. Their happiness degree prediction DNN was based on the answers to five psychometric questionnaires. As can be expected, their model performed much better than the traditional methodologies that assume a linear relationship between the variables. Since the deep neural network allowed for non-linear relationships to be estimated, it was able to better capture the complex structure or nature of happiness.\newline
The interrelationships of the factors affecting happiness have also been studied individually to a great extent in the past. Higher income has been associated with higher longevity or healthy life expectancy \cite{braveman2010socioeconomic}. However, the relationship between the GDP per capita of a country and its average healthy life expectancy is not as simple as it may seem at first glance. Granados et al.\cite{granados2008reversal} as well as Cutler et al.\cite{cutler2016economic}, showed a possibly negative correlation between the two.
Apart from this, previous studies [\cite{rasella2013impact}] have also explored the effects of income inequality (measured by the GINI household income in our data) with health and essentially life expectancy. \newline
Moreover, the average healthy life expectancy at birth is also shown to be affected by the available social support. Having better supportive, social relationships has been linked with better psychological regulation and thus better health and longevity [\cite{yang2016social}; \cite{ross2002family}].\newline
In addition to this, studies like the one by \cite{clausen2011corruption} have established a strong negative correlation between perceptions of corruption and confidence in public institutes or the national government. Further, confidence in the national government is shown to have a positive effect on the generosity of the residents of a country \cite{stagnaro2017good}.
Previous research has also shown that positive affect promotes generosity in individuals. \cite{moore1973affect} studied this effect in 7-8 year olds. In the book titled Handbook of Psychology \cite{isen2002role}, Isen confirmed the same.\newline
Additionally, Lucic \cite{luvcic2016causality} established a mutual, causal relationship between GDP and corruption. Experimental research \cite{nishi2015inequality} has also suggested that social support can be undermined by manifestations of inequality. Further, social support has been shown to reduce negative affect [\cite{wang_cai_qian_peng_2014}; \cite{cohen_wills_1985}].\newline
\newpage
Previous research \cite{kim_heshmati_2019} has also established the positive and robust effect of democracy (approximated by democratic quality) on GDP per capita. On the other hand, deliver quality (or government effectiveness) is negatively correlated with income inequality \cite{shafique_haq_2006} and corruption \cite{montes_paschoal_2015} and positively correlated with democratic quality \cite{magalhaes2014government}.\newline
Finally, several different research studies have shown that the factors that affect happiness include generosity \cite{park2017neural}, GINI index of household income or income inequality \cite{lang2012most}, GDP per capita \cite{abounoori2013macroeconomic}, healthy life expectancy at birth, social support \cite{gulaccti2010effect}, freedom to make choices, perceptions of corruption, confidence in the national government, positive affect and negative affect \cite{clark2017key} as well as delivery and democratic quality \cite{owen_videras_willemsen_2008}; World Happiness Report,2020 \cite{helliwell2020world}. All the aforementioned results have been summarized by a knowledge graph in Fig-1.\newline
In the recent past, feed-foward neural networks have found countless applications in different areas. Alas, even after their great approximating power, due to a large number of parameters and non-convex nature of the underlying loss functions they often suffer from the curse of getting stuck at local minimas. To prevent that,  F. Specht \cite{specht1991general} proposed a one pass learning algorithm General Regression Neural Network(GRNN) with a rigid structure. The network has found applications in learning the dynamics of a plant model for prediction or control F. Specht \cite{specht1991general}, Hong-Ze Li \cite{li2013hybrid} in  annual power load forecasting , Kamer Kayar \cite{kayaer2003medical} in medical diagnosis on Pima Indian Diabetes. Through our work we describe the ability of GRNNs to perform significantly better when compared to state of the art predictive models on a small and complex survey dataset.  \newline
In the next step, we discretize our dataset using a manual discretization scheme and combine knowledge by learning many Bayesian Networks(BNs) which serves as an expert system to generate interesting insights when queried. Our inspiration to use BNs on survey data in order to understand the relationships between the various variables is drawn from \cite{kitson2019learning} where the authors explore the factors that influence childhood mortality from preventable diseases and \cite{scutari2017bayesian} where malocclusion data is analyzed using BNs.

\begin{figure*}
  \includegraphics[width=1\textwidth]{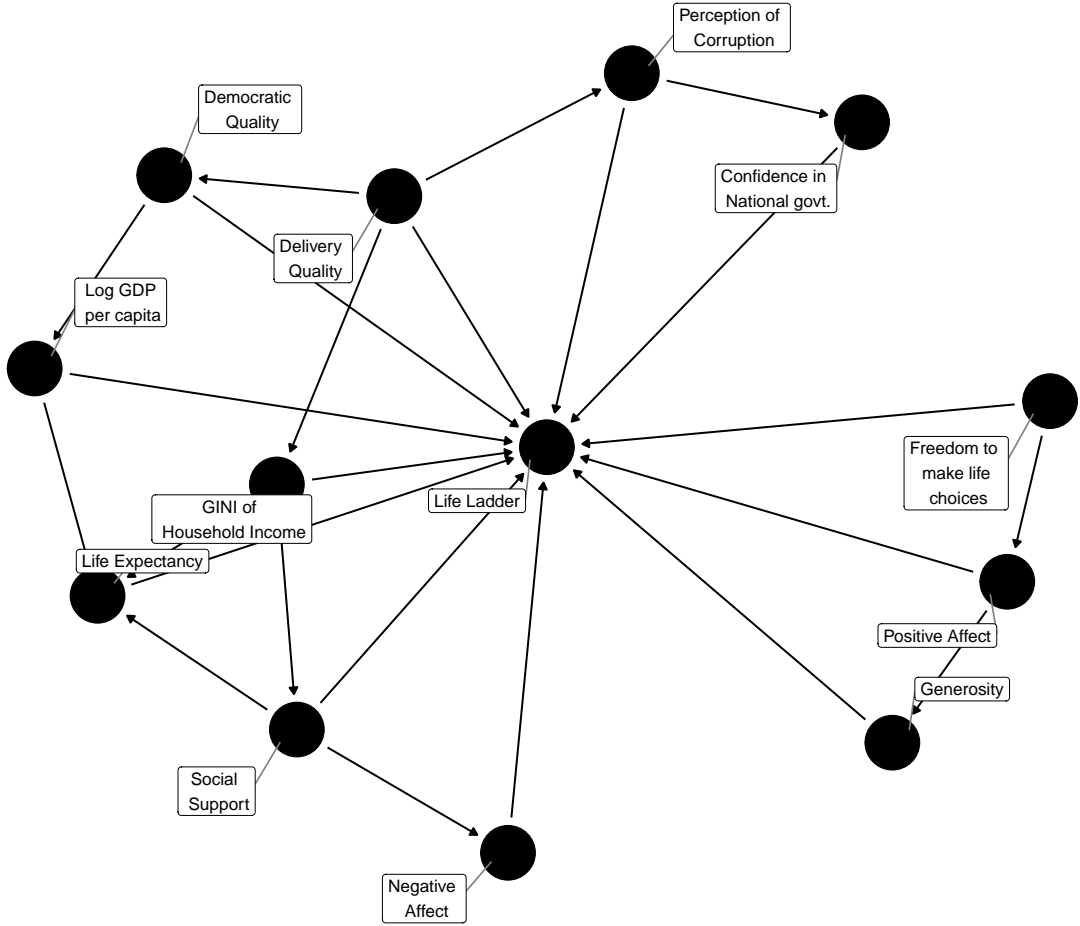}
\caption{A knowledge graph summarizing the past literature related to how the descriptors considered in the present study affect happiness.}
\end{figure*}

\section{Computational Methods}
\label{compmethods}
\subsection{Dataset Construction}
World Happiness Report is a landmark survey of the state of global happiness released every year since 2012 that ranks 156 countries by how happy their citizens perceive themselves to be. The report also contains online data about World Happiness which after some modifications was used by us to conduct the analysis.\newline
The World Happiness Report 2020 in turn gathers its data from the Gallup World Poll which contains survey data about 156 countries from 2005 to 2019 showing how the principal measures of happiness, and their main supporting factors, have evolved as a function of time. Even though the survey claims to have included data about 156 countries, there are only a small number of countries for which data has been recorded in the initial years.\newline
\textit{ Burundi, Jamaica, Somalia, Maldives, Trinidad and Tobago} 
have only started the survey process recently so they had data for just 2019.
Similarly, \textit{Congo (Kinshasa), Malaysia, Comoros, Central African Republic, South Sudan,Swaziland, Bahrain} had recorded data from 2018 and 2019.
To have uniformity in our dataset so as to generate meaningful insights and build robust models, the countries mentioned above were not included in the analysis.
It was further noticed that \textit{Haiti, Hong Kong S.A.R. of China, Sri-Lanka, Lesotho, Mozambique, Paraguay, Laos, North Cyprus, Iceland, Kuwait, Namibia, Czech Republic, Pakistan, Gambia, Israel, Russia, Iraq, Armenia} had no data for 2016.
But since the size of our dataset was already quite small, it was not justifiable to discard valuable information only because they had one year of information(single data point) missing. Therefore these countries having 3 years of data were still included. In the end, we had compiled information from last 4 years (i.e. 2016-19) comprising of 139 countries. A detailed description of all the features used for building models has been given in Table-2.\newline
Feature wide missing values were imputed by calculating the mean of the other year observations grouped by country for that specific feature. Features such as \textit{Democratic Quality} and \textit{Delivery Quality} had missing values for the year of 2019 for every country so they were replaced by the countrywise-means for 2016-2018. As a last resort, countries which had missing values for all four years for a particular feature were filled up using the mean across all countries. \newline
The target variable to be predicted was the countries \textit{Happiness Index} denoted by \textit{\textbf{Life Ladder}}. It has been described by the World Happiness Report, as the score calculated based on the average answers to the Cantril ladder life evaluation question in the Gallup World Poll. Respondents are asked to think of a ladder, with the best possible life for them being a 10, and the worst possible life being a 0. Then they are asked to rate their own current lives on that 0 to 10 scale.\newline
For the effective implementation and application of a BN to the dataset, it had to be discretized. Beuzen et. al [2018] gives a nice comparison of the different discretization schemes.  At first, a multi interval discretization scheme proposed by Fayad \& Irani [1993] was used to automatically generate bins for every continuous feature that offered the highest accuracy while predicting \textit{Life Ladder}. This method is categorized as a supervised disctretization method. However, the number of intervals returned by the same were too high to be fed into the conditional probability tables for the BN to be efficiently and realistically applied. So, a more manual approach which combined knowledge from the distribution of individual features and offered physical interpretibility was adopted. The number of intervals for each variable were limited to three, representing \textit{low, medium} and \textit{high}. The edges for each of the intervals were carefully chosen such that the length of the intervals wasn't too uneven and, at the same time, the frequency distributions of the intervals roughly represented the distribution of the continuous variables. All the variables with their respective bin ranges after the discretization procedure described above has been ellaborated in Table-1.

\begin{table}[hb]
\centering
\caption{Bin ranges obtained after the manual discretization procedure. The bins are chosen on the basis of frequency distributions and to maximize interpretability during inference later on.}
\begin{tabular}{llll}
\textbf{Variable} & \textbf{Low} & \textbf{Medium} & \textbf{High}  \\
\noalign{\smallskip}\hline\noalign{\smallskip}
 Log GDP per Capita & (6.81, 8.57) & (8.57, 9.94) & (9.94, 11.46)\\
 Gini of Household Income & (0.19, 0.38) & (0.38, 0.57) & (0.57, 0.85)\\
 Generosity & (-0.33, -0.09) & (-0.09, 0.19) & (0.19, 0.66)\\
 Positive Affect & (0.32, 0.62) & (0.62, 0.75) & (0.75, 0.92)\\
 Negative Affect  & (0.09, 0.25) & (0.25, 0.36) & (0.36, 0. 0.59)\\
 Perceptions of Corruption & (0.04, 0.51) & (0.51, 0.77) & (0.77, 0.97)\\
 Confidence in National Government & (0.07, 0.41) & (0.41, 0.66) & (0.66, 0.99)\\
 Healthy Life Expectancy  & (46.59,60.62) & (60.62,69.13) & (69.13,77.11)\\
 Democratic Quality & (-2.38, -0.92) & (-0.92, 0.32) & (0.32, 1.58)\\
 Delivery Quality  & (-1.93, -0.47) & (-0.47, 0.67) & (0.67, 2.10)\\
 Freedom to Make Life Choices & (0.30, 0.66) & (0.66, 0.82) & (0.82, 0.99)\\
 Social Support  & (0.41, 0.71) & (0.71, 0.85) & (0.85, 0.98)\\
 Life Ladder & (2.37, 4.83) & (4.83, 6.18) & (6.18, 7.86)\\
 
\noalign{\smallskip}\hline
\end{tabular}
\end{table}

\begin{table}[ht]
\fontsize{8.1}{10.1}\selectfont
\caption{A description of features affecting World Happiness which were considered while building models. The features have been grouped into 4 main categories which describe the \textit{Economic, Personal, National} and \textit{Social} aspects of a nation.}  
    \begin{tabularx}{\linewidth}{ l l X}
        \toprule
    \textbf{Category} & \textbf{Variable} &  \textbf{Description}    \\
        \midrule
\multirow{ 2}{*}{Economic} & Log GDP per Capita & The Statistics of GDP per capita in terms of Purchasing Power Parity (PPP) adjusted to constant 2011 international dollars, taken from the World Development Indicators (WDI)\\
& Gini of Household Income &  A measure of statistical dispersion intended to represent the income inequality or wealth inequality within various households of a nation or any other group of people.\\

\midrule
 \multirow{ 3}{*}{Personal} & Generosity & The residual of regressing the mean (national level) of the response to the survey question asking if an individual has donated money to a charity in the past month on GDP per capita\\
 & Positive Affect & The mean of 3 positive affect measures - happiness, laughter and enjoyment - in GWP (Gallup World Poll)\\
 & Negative Affect & The mean of 3 negative affect measures - worry, sadness and anger - in GWP\\
 
 \midrule
 \multirow{ 5}{*}{National} & Perceptions of Corruption & The national average to GWP survey responses to two questions asking whether or not widespread corruption exists in the Government and businesses respectively\\
  & Confidence in National Government & The national average of the response to a survey question by the GWP determining whether people have confidence in their national government\\
  & Healthy Life Expectancy & The healthy life expectancy at birth data extracted from the World Health Organization’s (WHO) Global Health Observatory data repository (interpolation and extrapolation used to meet time frame of world happiness report)\\
  & Democratic Quality & A measure of people's access to power through factors like freedom of association, political stability and ability to participate in selection of government\\
  & Delivery Quality & A measure related to the exercise of power including government effectiveness, rule of law and controlling corruption\\
\midrule
\multirow{ 2}{*}{Social} & Freedom to Make Life Choices & The national mean of the answer to the GWP survey question asking if a person is satisfied or dissatisfied with their freedom to choose what they do with their life\\
  & Social Support & The national average of the responses to a GWP survey question determining whether a person has someone to count on in times of trouble\\
\midrule
    \end{tabularx}
\end{table}

\subsection{Model Building}
\subsubsection{Predictive Models}
Data from 2016-18 was used to train the predictive models and their performance was evaluated on 2019 data. Initially when an Ordinary Least Squares(OLS) model was learned on top of the features described in Table-2, it achieved an $R^2$ of 0.75 on the Test Data. The $R^2$ value suggests that a Linear Model moderately explains the variance between Life Ladder and other features.\newline
A variety of different ML models such as(but not limited to) Regularized Linear (Ridge, Lasso, Elastic Net), Tree and Ensemble models (Decision Trees, Gradient Boosted Trees, Random Forests) and Deep Learning models (Shallow Neural Networks with 2 hidden layers and different combinations of neurons as well as a more state-of-the-art architecture using techniques like batch normalization and adaptive learning rates using the fastai framework) when learned on our dataset, gave mostly moderately accurate results with the state-of-the-art DNN performing the best.
 It was then that we tried out a radial basis function network model, more commonly known as General Regression Neural Network (GRNN). F. Specht has \cite{specht1991general} claimed that for problems wherein the assumption of Linearity is not satisfied, GRNNs, a one pass learning algorithm with highly parallel structure seems to work well. In the past, GRNNs have been used for Adaptive Control Systems and in building models on patient’s medical data to study obesity \cite{garaigordobil2015predictor}.\newline
GRNNs are used to estimate linear or nonlinear regression where the dependent variable is continuous. They are useful for regression modeling problems because they have strong non-linear mapping capability and can be computed relatively quickly. Robust to outliers they can solve any function approximation problem if sufficient data is provided. In addition, they tend to perform well in terms of prediction accuracy, even with small samples. So, given we only had about 400 data points to train models on, using GRNNs seemed to be a viable option.\newline
GRNNs have four layers, namely an input layer, a pattern layer, a summation layer (which contains two neurons) and an output or decision layer. The input layer contains one node for each attribute so our model contained 12 input nodes. It serves to distribute the input data to the pattern layer.
The pattern layer contains one node for each training case. For a sample consisting of about 400 training data points there will be about 400 pattern layer nodes.  The activation function of pattern layer neuron i is: 
$P_{i} = e^{D^{2}/2\sigma^{2}}$ \newline
Where $D^2 = (X-X_{i})^{T} (X-X_{i})$ is the squared euclidean distance between the input vector X and the $i^{th}$training input vector $X_i$; and $\sigma$ determines the spread/shape of the distribution. It is known as the smoothing factor. Greater the value of smoothing factor, the more significant distant training cases become for the predicted value. The larger the smoothing value the smoother the function is, and in the limit, it becomes a multivariate Gaussian. Smaller values of the smoothing parameter allow the estimated density to assume non-Gaussian shapes. The best value of this parameter was found using cross validation on our data. The output of the neuron is therefore a measure of the distance of the input from the observed patterns. The computed values are then passed to the summation layer nodes. Notice that each pattern layer neuron is connected to two neurons in the summation layer.\newline

The summation layer has two types of summation neurons. 
The first node($S_N$), computes the sum of the weighted outputs of the pattern layer: $S_{N}=\sum^{n}y_{i}P_{i}$ and the second ($S_{D}$) calculates the unweighted outputs of the pattern neurons: $S_{D} = \sum^{n}P_{i}$ . The Output node performs a normalization by dividing the output from the summation layer to generate the prediction: $Y^{\wedge}(X) = S_{N}/S_{D}$ . The complete architecture of the model has been visualized in Fig-2.\newline
While the modelling process followed above gives us sufficiently accurate predictive power for forecasting Happiness for a given country, it lacks in terms of interpretability and addressing the uncertainity involved with survey data. Now, in order to generate some interpretations, we went on to build a Bayesian Network on top of the features described in Table-1.

\begin{figure*}
  \includegraphics[width=0.9\textwidth]{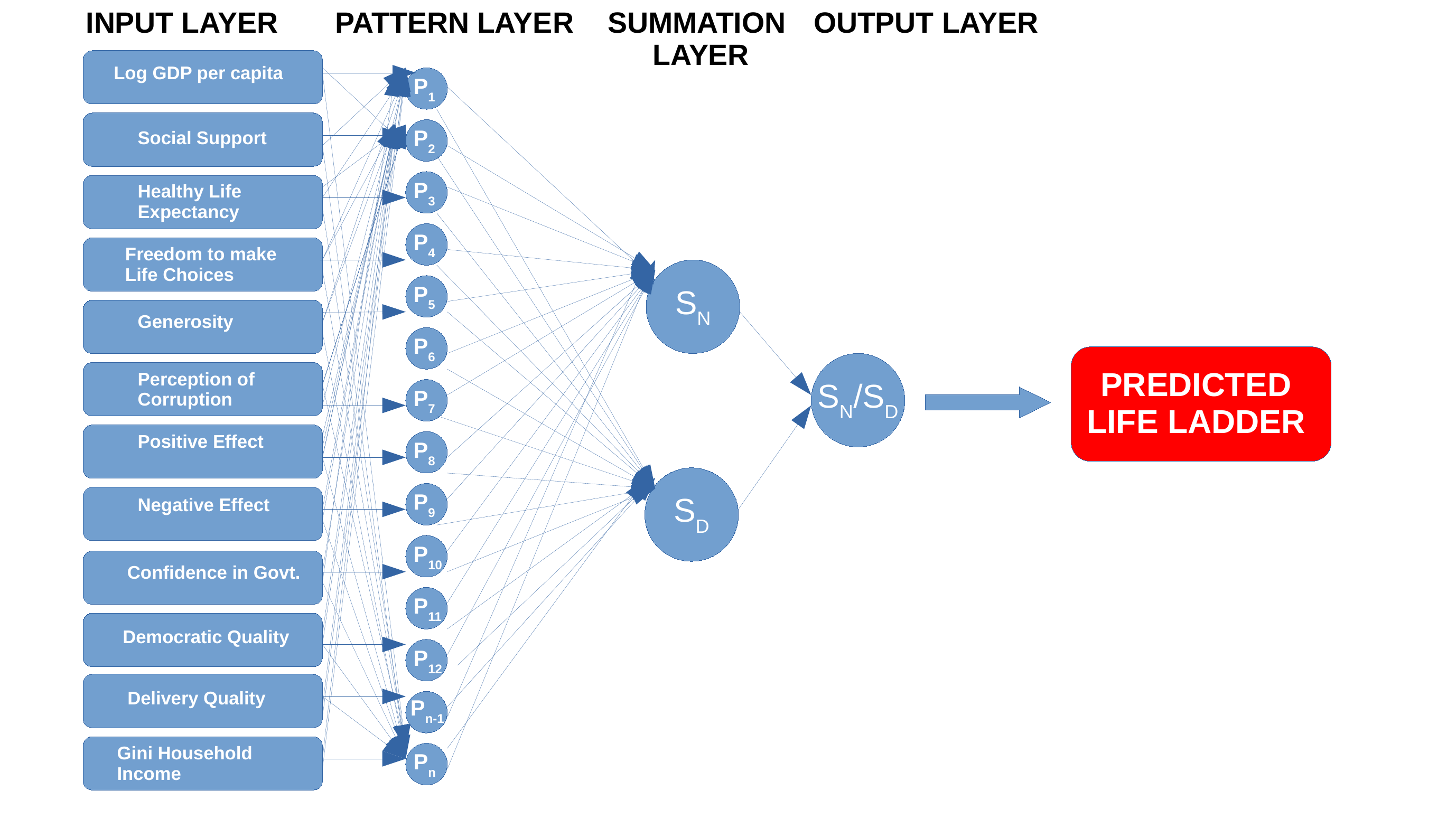}
\caption{Describes the model architecture used for predicting Life Ladder.}
\end{figure*}

\subsubsection{Bayesian Networks}
Bayesian Networks(BNs) allow us to discover causal structures in raw statistical data. First developed in the late 1970’s, they are directed acyclic graphs (DAGs) in which the nodes represent variables of interest and the links represent informational or causal dependencies among the variables. The strength of a dependency is represented by the conditional probabilities that are attached to each cluster of parents-child nodes in the network \cite{koller2009probabilistic}. \newline
The graphical structure $G = (V,E)$ of a BN is a DAG,
where $V$ is the node/vertex set and $E$ is the edge set. The DAG defines a
factorization of the joint probability distribution of $V = \{X_{1}, X_{2}, . . . , X_{v}\}$, often called the global probability distribution, into a set of local probability distributions, one for each random variable.
The form of the factorization is given by the Markov property of Bayesian networks \cite{korb2010bayesian}, which states that every random variable $X_i$ directly depends
only on its parents $\Pi_{X_{i}}$. Moreover, since we had already discretized all the continuous features in our World happiness dataset, the joint probability distribution assumed the following form :
\begin{equation}
    P(X_{1},X_{2},...,X_{V}) = \Pi_{i=1}^{v} P(X_{i}|\Pi_{X_{i}})
\end{equation}
\newpage
In the context of social sciences, the structure of the DAG may identify which nodes are directly related to the target of the analysis and may therefore be used to improve the process of devising policies keeping the happiness of a country in mind. The Bayesian Network on World Happiness data was learned using the Hill Climb Algorithm which is a greedy method for optimization. This learning approach is computationally efficient and has been found to provide very good results in most cases \cite{tsamardinos2006max}. Using bootstrapping, 10000 different BNs were learned using the hill climb algorithm while keeping Bayesian Information Criteria (BIC) as the evaluation metric. A consensus model was obtained by averaging predictions and keeping arcs which occurred in at least 
50\% of the networks.

\section{Results \& Discussions}

A variety of different predictive models were tried out in order to predict the \textit{Life Ladder} of a country on the basis of variables described in Table-1. A cohort of over 30 different machine learning models, were trained on data from 2016-18 and their performance was evaluated on the data available for 2019. The performance of some standard models along with the best performing model has been summarized in Table-3. \newline
The models were judged and compared using 3 metrics viz. $R^2$, \textit{MAE, MSE} while forecasting \textit{Life Ladder} values for different countries for 2019. The $R^2$ value is a statistical measure that represents the proportion of the variance for \textit{Life Ladder} that is explained by independent variables in a regression model. Even though having high $R^2$ value is a necessary condition to judge a models performance on unseen data, it is not a sufficient one. 
So to be completely sure that our chosen model would perform well on unseen data, the mean absolute error \textit{(MAE)} and mean squared error \textit{(MSE)} were also calculated for the model predictions.\newline
It was observed that GRNNs gave the best performance when compared to its counterparts, obtaining an $R^{2}$=0.88, \textit{MAE}= 0.29 and \textit{MSE} of 0.15 respectively; followed by the DNN with $R^{2}$=0.86, \textit{MAE}= 0.29 and \textit{MSE} of 0.16 respectively. Popular Tree based ensemble models such Random Forests and Gradient Boosted trees also performed well giving $R^{2}$=0.84, \textit{MAE}= 0.32 and \textit{MSE} of 0.19; $R^{2}$=0.82, \textit{MAE}= 0.35 and \textit{MSE} of 0.23 respectively, while predicting Life Ladder for 2019.\newline
The results have been further investigated in Fig-3, which shows how the GRNN forecasts for Life Ladder compare with the actual life ladder values for the year 2019. Even with its relatively simple and rigid structure, the GRNN was able to predict the life ladder values for 2019 fairly accurately for most cases. The deviation of the predicted values from actual values in some cases could be attributed unknown factors other than those represented in our data. For instance, a study \cite{sciencedaily_2016} has shown that even genes may contribute towards making certain nations happier than the others. Interestingly, deviations from predicted values seem to be more concentrated in the lower half of life ladder values.\newline

\begin{figure*}[hb]
  \includegraphics[width=0.8\textwidth]{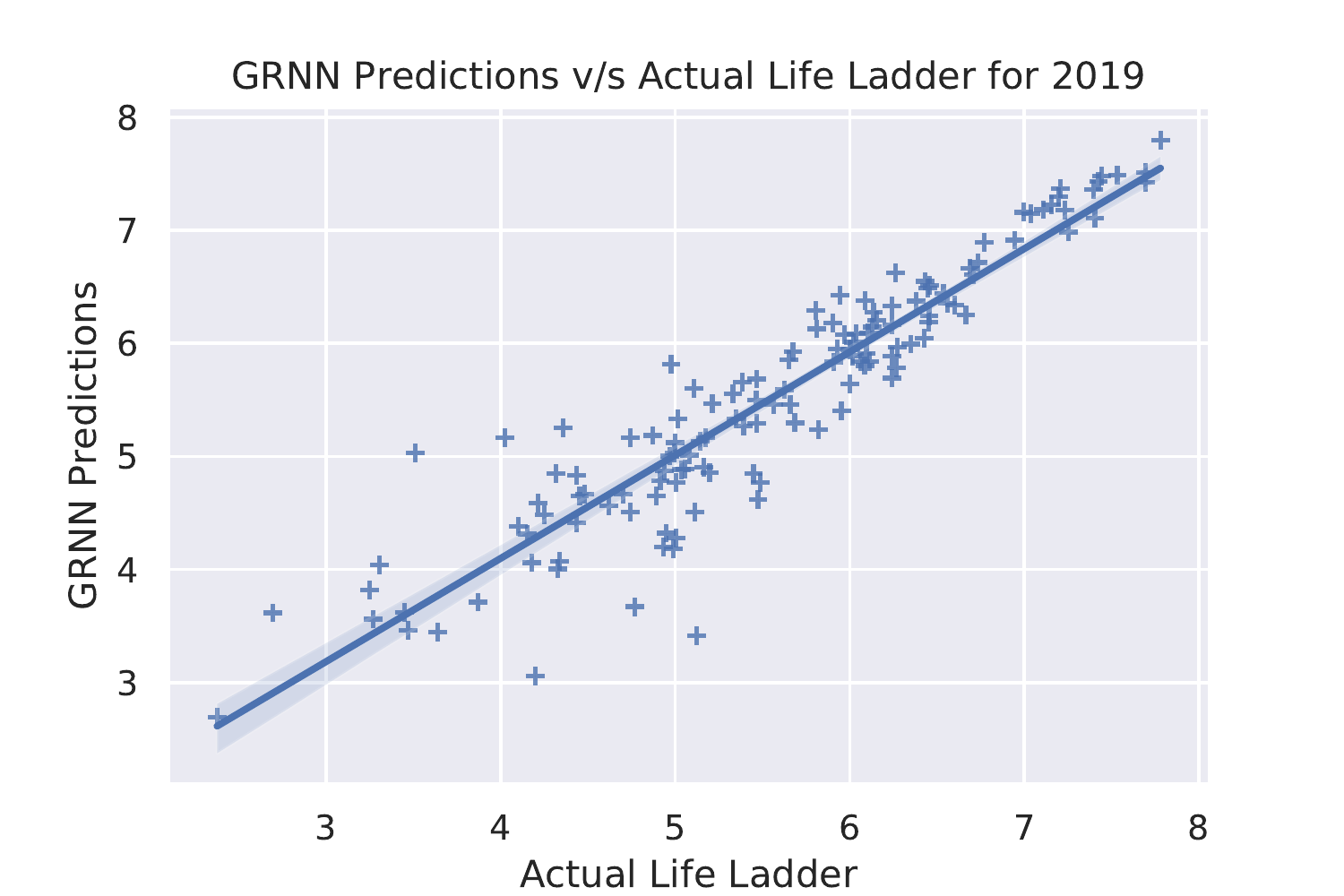}
\caption{GRNN predictions for the year 2019 and their comparison with actual data.}
\end{figure*}

\newpage
\begin{table}[ht]
\centering
\caption{Assessment of the performance of different Predictive Models on unseen data. GRNN obtains the highest $R^2$ as well as lowest MAE and MSE while predicting Life Ladder for 2019.}
\begin{tabular}{llll}
\hline\noalign{\smallskip}
Model & $R^2$ & MAE & MSE  \\
\noalign{\smallskip}\hline\noalign{\smallskip}
 GRNN & 0.88 & 0.29 & 0.15\\
 Deep NN & 0.86 & 0.29 & 0.16\\
 Random Forests & 0.84 & 0.32 & 0.19\\
 XGBoost & 0.82 & 0.35 & 0.23\\
 OLS  & 0.75 & 0.43 & 0.31\\
 Ridge & 0.75 & 0.43 & 0.31\\
 Decision Trees & 0.62 & 0.46 & 0.48\\
\noalign{\smallskip}\hline
\end{tabular}
\end{table}

The one major purpose behind studying human well-being and the factors affecting it has largely been centered around gathering valuable information that could help governments around the world maximize the well-being or life satisfaction of their citizens. With the same, purpose, we used the data from the World Happiness report to build a Bayesian Network (Fig. 4).\newline
Paying close attention how the various factors of happiness affect each other (especially how government and economic factors can affect personal and social factors) may be beneficial when it comes to policy making to maximize the well-being of people. The Bayesian Network obtained from the data can be used for the same.\newline
We observe that several relationships obtained in the network are as expected and backed up by previous research (depicted in the knowledge graph). As mentioned in the introduction section, GDP per capita is shown to influence the Healthy Life Expectancy of a nation. Similarly, Delivery Quality affects Democratic Quality and corruption; freedom to make life choices influences negative affect while social support influences negative affect.  The direction of the predicted relationship in some cases, however, was reversed. For instance, according to the bayesian network, GINI of household income is affected by the healthy life expectancy. However, the expected causal relationship between the two would be the opposite. Upon digging deeper, we found a paper \cite{hu_lenthe_mackenbach_2015} proving that in the European context, income inequality has little to no independent effect on the healthy life expectancy of a nation. Such a possible inconsistency in our data could be the reason for this reversal.\newline

\begin{figure*}[hb]
  \includegraphics[width=0.925\textwidth]{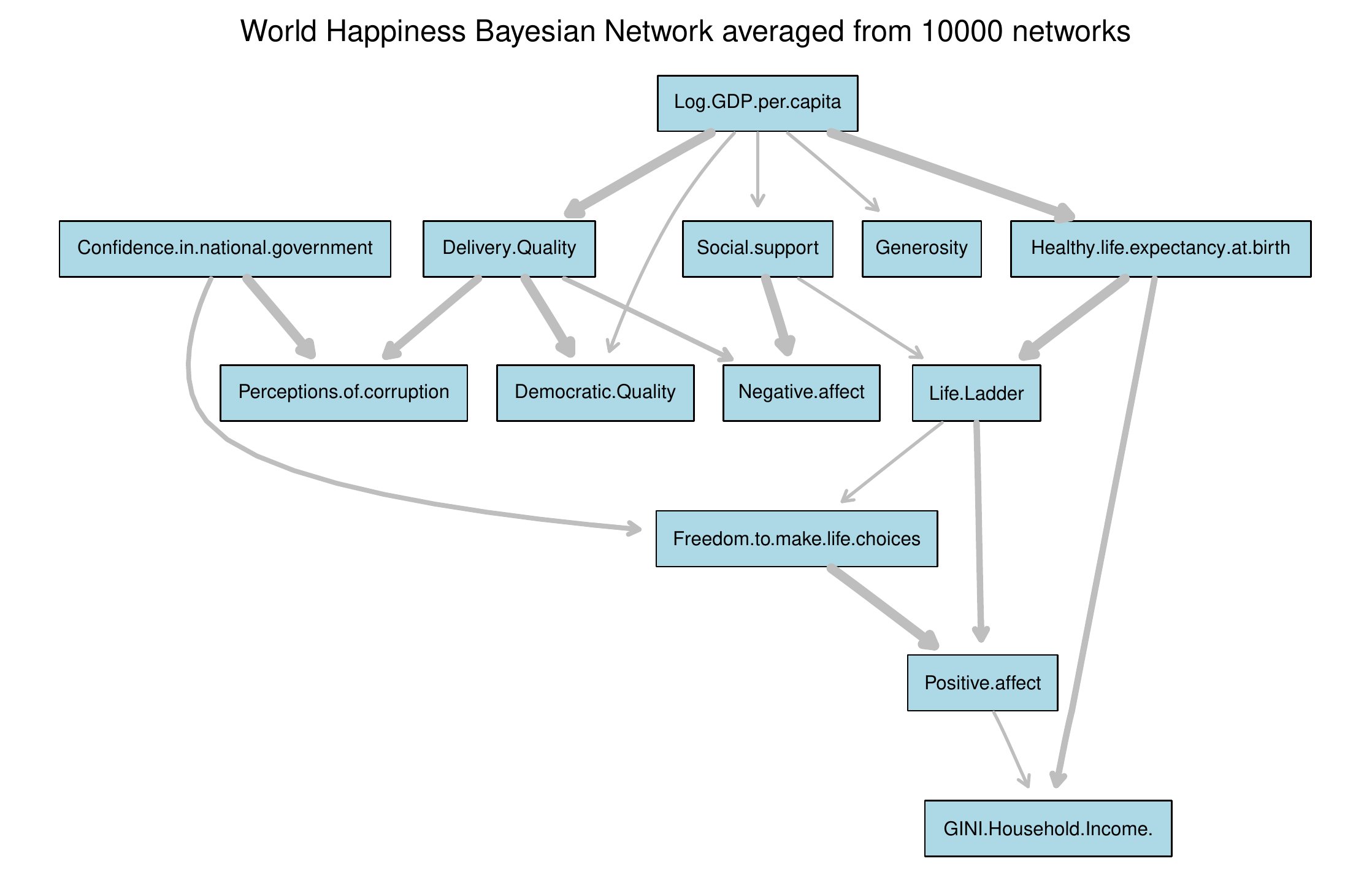}
\caption{The consensus model formed by bootstrapping and learning 10000 different Bayesian Networks. Only those arcs have been included which occur more than 50\% of the times. The arc thickness signifies their respective strengths.}
\end{figure*}

\newpage
The GDP per capita, in particular seems to have an influence on a lot of the variables from the World Happiness Report data. This, as mentioned in the knowledge graph (Fig-1), includes Healthy Life Expectancy. A close inspection in Fig-5 tells us that higher GDP per capita supports higher life expectancy. In fact, the probability of having Low healthy life expectancy(47-61 years) decreases sharply from 0.78 (given that the country has a low GDP per capita) to almost 0 when the country has High GDP per capita. For countries having moderate (close to average) GDP per capita values, their healthy life expectancy lies between 61-69 years with probability 0.78. 
Additionally, it can be said with 70\% certainty that a high GDP per capita will lead to a high healthy life expectancy (69-77 years) value for the country.

\begin{figure*}[ht]
\includegraphics[width=0.9\textwidth]{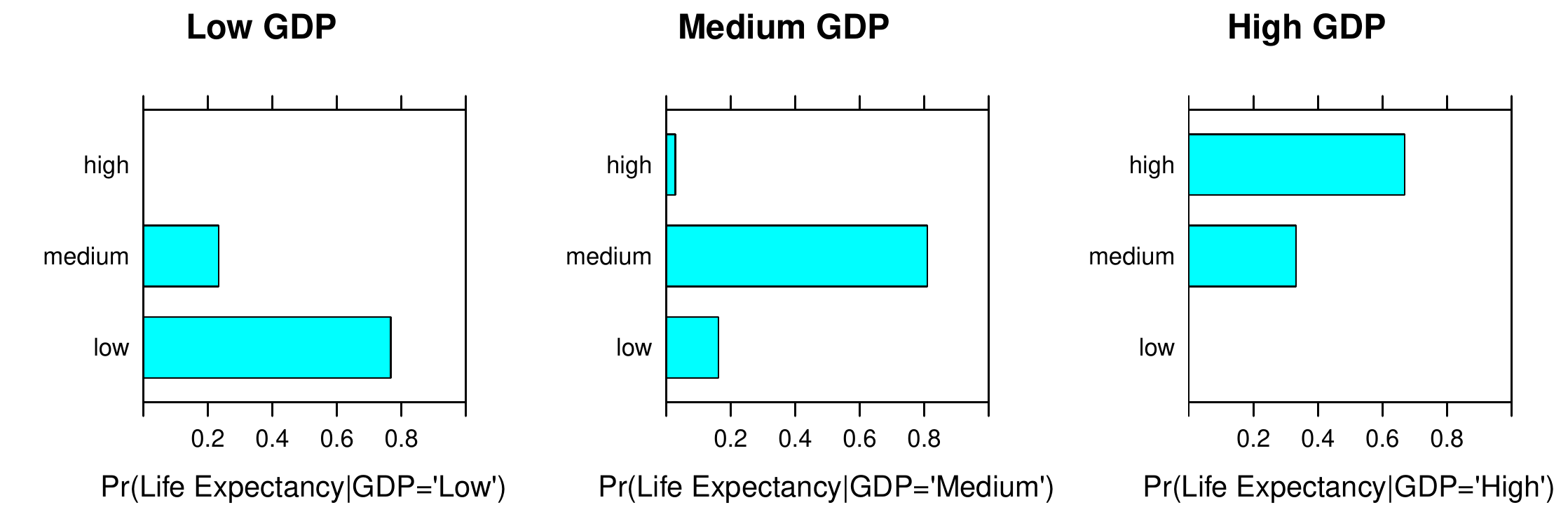}
\caption{Results of querying the BN to observe Healthy Life Expectancy given different states of GDP}
\end{figure*}
In addition to this, even though Generosity is calculated as the residual of regressing national average of response to the GWP question “Have you donated money to a charity in the past month?” on GDP per capita, the data seems to suggest a link between GDP per capita and generosity. A closer inspection of the conditional probabilities related to the same reveals mostly moderate to low generosity levels, which might be attributed to the weak link from GDP to Generosity in the learned BN (Fig-4). It is also evident from Fig-6 that people from Low GDP countries seem to be donating more than medium and almost the same as high GDP countries.\newline
\begin{figure*}[hb]
\includegraphics[width=0.9\textwidth]{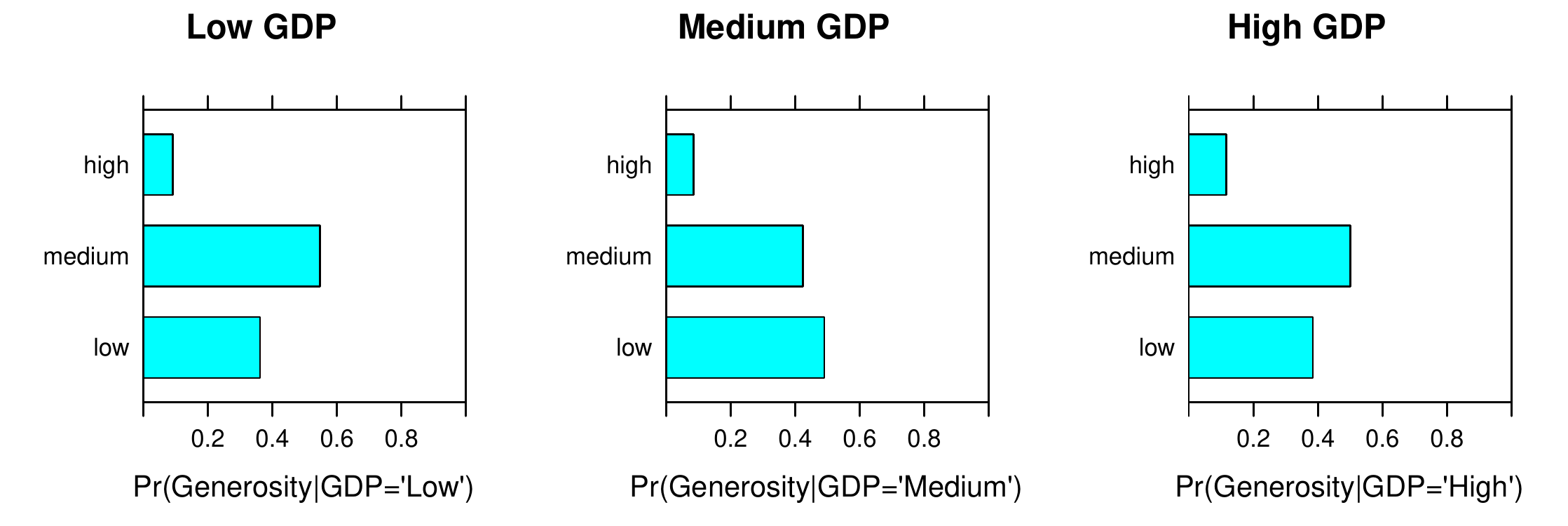}
\caption{Results of querying the BN to observe Generosity given different states of GDP}
\end{figure*}

The network also indicates that the Delivery quality of a government is affected by the GDP per capita of the country and then goes on to strongly influence the perceived corruption in the nation. This mediated relationship between the GDP per capita and perception of corruption can be used to explain the mutual, causal relationship between GDP and corruption established by previous research \cite{luvcic2016causality}. Since increase GDP positively impacts Delivery quality which in turn would reduce corruption, we would expect GDP to be negatively correlated with perception of corruption. The same is confirmed by Fig. 7 below. Notice how the perceptions of corruption amongst people has a 63\% chance of being high if the GDP of that nation is Low. On the other hand, the chance of having a low perception of corruption gradually increases from 1\%(for Low GDP countries) to 34\% for high GDP countries.\newline

\begin{figure*}[ht]
\includegraphics[width=0.9\textwidth]{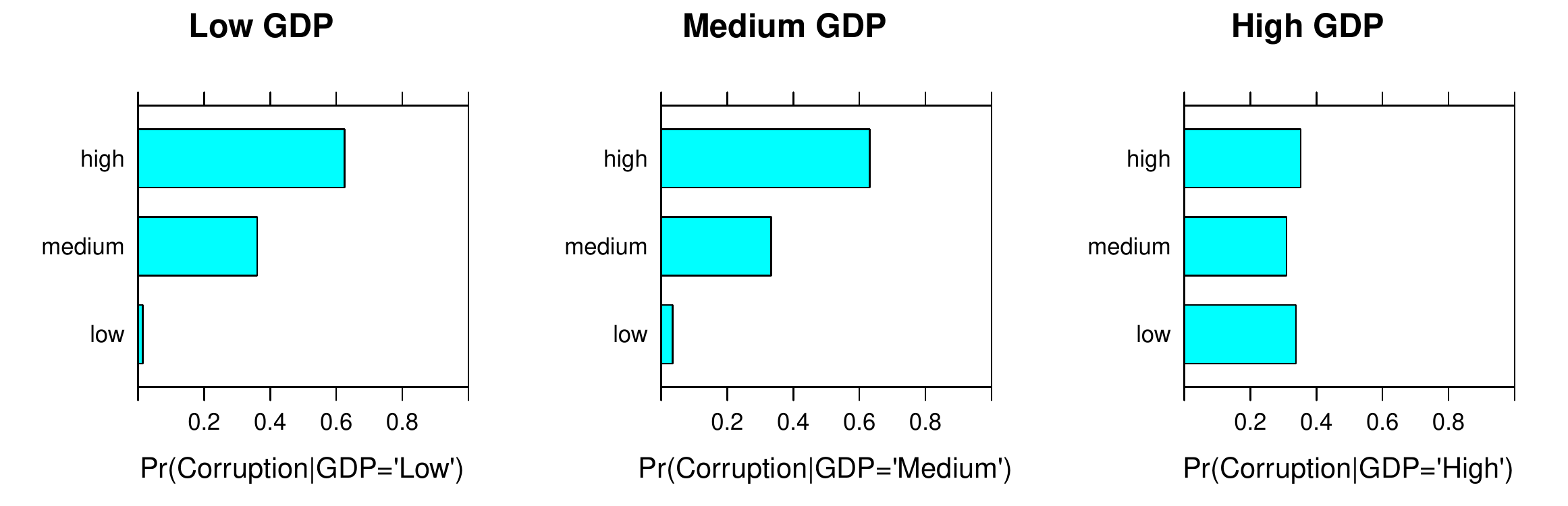}
\caption{Results of querying the BN to observe perceptions of corruption given different states of GDP}
\end{figure*}
\newpage
Not only this, GDP per capita also seems to have an impact social support. However, despite the apparent influence of GDP per capita on a number of the factors affecting happiness, research strongly suggests that more money does not always mean more happiness. In fact, beyond a certain threshold, GDP per capita no longer holds even a short term positive correlation with happiness \cite{easterlin_2014}.\newline
Another rather interesting causal relationship that came up in the Bayesian Network is the apparent influence of confidence in national government on an individual's perceived freedom to make life choices. Fig-8 explores the same, plotting the probabilities associated with a given amount of freedom available to make life decisions conditioned on the varying levels of confidence people of a given nation have in their government. It is observed that the probability of having Low freedom to make life choices reduces from 0.37 to 0.01 as the given confidence in government transitions from Low to High. As can be expected, the probability of having High freedom to make life choices increases from 0.30 to 0.68 as the given confidence in government transitions from Low to High.

\begin{figure*}[hb]
\includegraphics[width=0.9\textwidth]{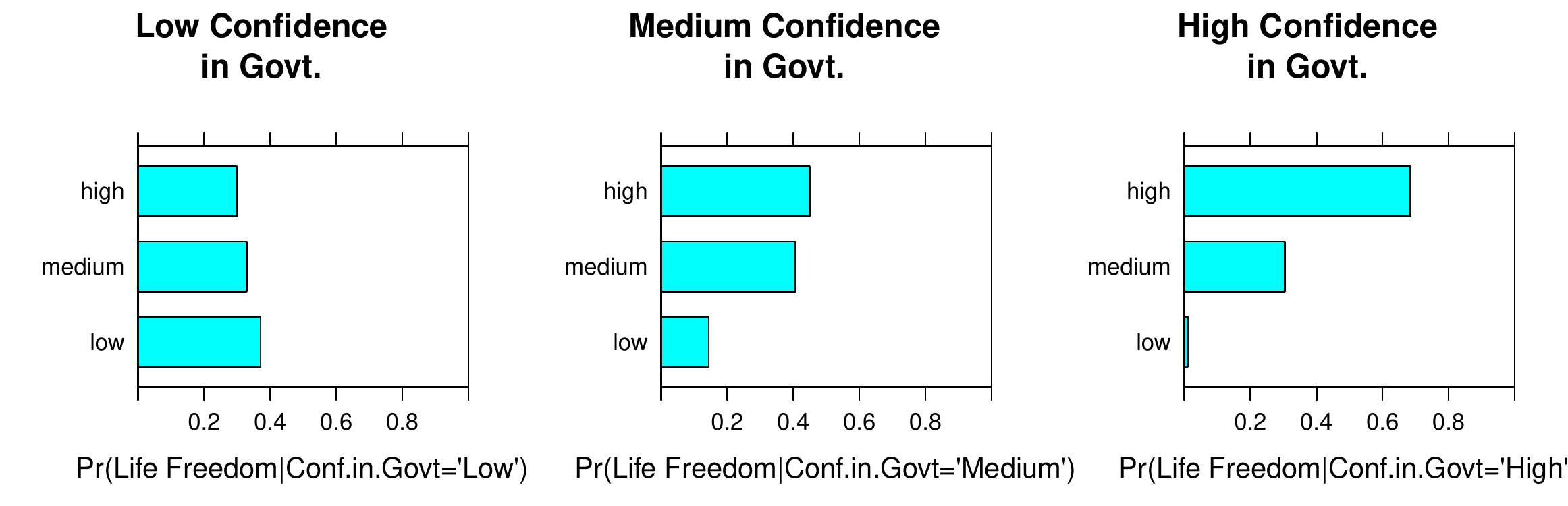}
\caption{Results of querying the BN to observe Freedom people of a given nation have to make life choices given different levels of confidence in their national government.}
\end{figure*}

\section{Conclusions}
In this study, we conducted an analysis of the World Happiness data of 156 countries. A knowledge graph representing past literature related to Happiness and the factors affecting it was created in accordance with the variables present in our data. After appropriate pre-processing, resultant data was analyzed using 2 types of computational strategies viz. \textit{Predictive models}: for predicting the happiness index(\textit{‘Life Ladder’}) of a country and \textit{Bayesian Networks}: for exploring causal relationships among variables.
Over 30 different Machine Learning and Deep Learning models were trained on happiness data from 2016-18 and their performance was evaluated while forecasting Happiness Index for 2019. GRNN, a one pass learning neural network with a highly parallel structure outperformed other state of the art models offering \textit{MAE} as low as 0.29 on unseen data of 2019. For the effective learning (and later on inference) from Bayesian Networks, a manual discretization scheme based on frequency distributions and maximizing interpretability was used to discretize continuous random variables into discrete ones with 3 levels viz. \textit{Low}, \textit{Medium} and \textit{High}. The greedy hill climb algorithm was used to learn over 10000 different BN structures using bootstrapping and an averaged World Happiness network was fixed. To query this network, exact inference through a number of conditional probability queries amongst the features affecting happiness was performed. The major goal behind the construction and querying of the aforementioned Bayesian Network was to enable a better and more comprehensive understanding of the factors affecting happiness and the relationships between them. Apart from the expected links among the variables, a few unexpected associations were also uncovered and may be worth studying further. One such relationship was the one between the GDP per capita and generosity of a nation. Contrary to expectation, generosity did not simply increase with the increase in GDP per capita. Another new and interesting possible relationship revealed was the one between confidence in national government and freedom to make life choices. To the best of our knowledge, no study so far has talked about a link between these two variables. It could be of interest to investigate the existence of the same more deeply. Not only was the BN able to bring forward interesting interrelationships of the factors affecting happiness, it could also be used to give direct, logical explanations to already established relationships like that between GDP and perception of corruption. Any knowledge gained about happiness and the factors affecting it can be invaluable in the process of holistic policy-making. This study is an attempt to contribute to the same.

\section*{Acknowledgements}
We would like to thank the Department of Mathematics at Shiv Nadar University for extending support throughout the duration of this project.
\section*{Declarations}
\textbf{Funding} \newline The author(s) received no financial support for the research and authorship of this article. \newline 

\textbf{Compliance with ethical standards}\newline Conflict of interest On behalf of all authors, the corresponding author states that there is no conflict of interest.\newline

\textbf{Availability of data and material} \newline Data used in the present study is openly available at \textit{\href{https://worldhappiness.report/ed/2020/}{World Happiness Report 2020}}\newline 

\textbf{Code availability} All the code to reproduce the results can be found on the github repository for this project:\newline \textit{\href{https://github.com/Sid-darthvader/Network_Learning_approaches_to_study_World_Happiness/}{https://github.com/Sid-darthvader/Network\_Learning\_approaches\_to\_study\_World\_Happiness/}}

\small\bibliographystyle{plain}

\end{document}